\documentclass[11pt,a4paper,fleqn]{article}
\catcode`\@=11
\@addtoreset{equation}{section}
\def\theequation{\arabic{section}.\arabic{equation}}
\def\appendix{\renewcommand{\thesection}{\Alph{section}}\setcounter{section}{0}
              \renewcommand{\theequation}
            {\mbox{\Alph{section}.\arabic{equation}}}\setcounter{equation}{0}}

\def\maketitle{\thispagestyle{empty}\setcounter{page}0\newpage
                \renewcommand{\thefootnote}{\arabic{footnote}}
                  \setcounter{footnote}0}
\renewcommand{\thanks}[1]{\renewcommand{\thefootnote}{\fnsymbol{footnote}}
               \footnote{#1}\renewcommand{\thefootnote}{\arabic{footnote}}}

\renewcommand{\title}[1]{\begin{center}\Large\bf #1\end{center}\rm\par\bigskip}

\renewcommand{\author}[1]{\begin{center}\Large #1\end{center}}
\newcommand{\address}[1]{\begin{center}\large #1\end{center}}

\def\babs{\hrule\par\begin{description}\item{Abstract: }\it}
\def\eabs{\par\end{description}\hrule\par\medskip\rm}
\renewcommand{\date}[1]{\par\bigskip\par\sl\hfill #1\par\medskip\par\rm}
\newcommand{\ack}[1]{\par\section*{Acknowledgments} #1}

\def\segue{\qquad\Longrightarrow\qquad} 
\def\hs{\qquad}               
\def\nn{\nonumber}            
\def\beq{\begin{eqnarray}}    
\def\be{\begin{eqnarray}}
\def\eeq{\end{eqnarray}}      
\def\ee{\end{eqnarrayn}}
\def\ap{\left.}               
\def\at{\left(}               
\def\aq{\left[}               
\def\ag{\left\{}              
\def\cp{\right.}              
\def\ct{\right)}              
\def\cq{\right]}              
\def\cg{\right\}}             

\def\R{{\hbox{{\rm I}\kern-.2em\hbox{\rm R}}}}   
\def\H{{\hbox{{\rm I}\kern-.2em\hbox{\rm H}}}}   
\def\N{{\hbox{{\rm I}\kern-.2em\hbox{\rm N}}}}   
\def\C{{\ \hbox{{\rm I}\kern-.6em\hbox{\bf C}}}} 
\def\Z{{\hbox{{\rm Z}\kern-.4em\hbox{\rm Z}}}}   
\def\ii{\infty}                                  
\def\X{\times\,}                                 
\def\Tr{\mathop{\rm Tr}\nolimits}                  
\renewcommand{\Re}{\mathop{\rm Re}\nolimits}       
\def\dir{/\kern-.7em D\,}                          
\def\lap{\Delta\,}                                 
\def\al{\alpha}
\def\ga{\gamma}
\def\de{\delta}
\def\ep{\varepsilon}
\def\ze{\zeta}

\def\la{\lambda}

\def\si{\sigma}
\def\om{\omega}

\def\Ga{\Gamma}

\def\La{\Lambda}

\def\be{\begin{equation}}
\def\ee{\end{equation}}
\def\bea{\begin{eqnarray}}
\def\eea{\end{eqnarray}}

\def\nn{\nonumber}

\renewcommand{\title}[1]{\begin{center}\Large\bf #1\end{center}\rm\par\bigskip}
\renewcommand{\author}[1]{\begin{center}\Large #1\end{center}}

\begin{document}

\title{One-loop $F(R,P,Q)$ gravity in de Sitter universe}
\author{Guido Cognola\thanks{cognola@science.unitn.it}
and
Sergio Zerbini\thanks{zerbini@science.unitn.it}
}
\address{
Dipartimento di Fisica, Universit\`a di Trento \\
and Istituto Nazionale di Fisica Nucleare \\
Gruppo Collegato di Trento, Italia
}

\begin{abstract}
Motivated by the dark energy issue, the one-loop
quantization approach for a class of relativistic higher order
theories is discussed in some detail. 
A specific $F(R,P,Q)$ gravity model at the one-loop level in a de Sitter
universe is investigated, extending the similar program developed 
for the case of $F(R)$ gravity. The stability conditions under arbitrary
perturbations are derived,
\end{abstract}

\maketitle

\section{Introduction}

It is well known that recent astrophysical data indicate that our universe is currently in a  
phase of accelerated expansion. This is one of the most important achievement in cosmology. 
The origin of this observation is substantially not completely understood and the related issue is 
called the dark energy problem.

Several possible explanations have been proposed in literature, among them one of the most popular is
based on the use of  gravitational modified models, the simplest one being Einstein gravity plus the inclusion of a 
small and positive cosmological constant, and this model works quite well but, having however some drawbacks 
(see, for example \cite{review,fara10,toni} and reference therein).

Roughly, the idea is that Einstein gravity is only an approximate low energy contribution, and 
additional terms depending on quadratic curvature invariants should be included. The idea is quite old, one of the 
first proposal was contained in \cite{staro}, where quantum $R^2$ gravity modifications were investigated 
(for a review, see \cite{buch}). The inclusion of  general higher order contributions is also important for another aspect,
since, sometimes they give extra terms which may also realize the early time inflation \cite{seba08}.

In  previous papers \cite{cogno,guido2,cogno8,cogno9}, $f(R)$ gravity models and a non local Gauss-Bonnet gravity model  at the one-loop level in a de Sitter
background have been investigated. A similar program for the case of  pure Einstein gravity was initiated in refs. 
\cite{perry,duff80,frad} (see also \cite{ds1,vass92}). Furthermore, such approach
 also suggests  a possible way of investigating the cosmological constant
issue \cite{frad}. Hence, the study of one-loop generalized modified gravity  is a
natural step to be undertaken for the completion of such a program,
keeping always in mind, however, that a consistent quantum gravity theory is not available yet. 

Making use of generalized zeta-functions regularization (see, for
instance \cite{dowker,hawking,eli94,byts96,kirsten00}), one may evaluate the one-loop effective
action and then study the possibility of stabilization of the
de Sitter background by quantum effects. Recall that in the one-loop approximation, the theory can be conveniently
described by the (Euclidean) one-loop partition function (see  \cite{buch}). 
For example, in the simplest case of a scalar field, one has
 \beq Z=e^{-I[\phi_c]}\int
D\phi\:\:e^{-\int\,dV\,\phi L\phi}=e^{-\Gamma[\phi_c]}\,. 
\eeq 
Here $I[\phi_c]$ is the classical action, evaluated on the background field  $\phi_c$, while $\Gamma$
is the one-loop effective action, which can be related to the
determinant of the fluctuation operator $L$ by 
\beq 
\Gamma=-\ln
Z=I+\frac{1}{2}\ln\det\frac{L}{\mu^2}\,, \label{I} 
\eeq 
$\mu^2$ being a renormalization parameter, which appears for dimensional
reasons. Of course, in dealing with  gauge theories, one needs a gauge braking term and the related F-P ghost contribution. 

The functional determinant may formally expressed by 
\beq \ln \det
\frac{L}{\mu^2}=-\int_0^\ii dt\ t^{-1} \Tr e^{-tL/\mu^2}\,.
\label{d} \eeq 
Here the  heat trace $\Tr e^{-tL}$ plays a preeminent
role. In fact, for a second-order, elliptic non negative differential operator
$L$ in a boundaryless {\it compact} d-dimensional  manifold, one has
the small-$t$ asymptotic heat trace expansion 
\beq \Tr
e^{-tL}\simeq\sum_{j=0}^\ii A_j(L) t^{j-d/2}\:, \label{tas00} \eeq
where $A_j(L)$ are the Seeley-deWitt coefficients
\cite{dewi65b,seel67-10-172}. 
As a result, the expression (\ref{d}) is divergent and a regularization and renormalization are required.
Zeta-function regularization may be implemented by  \cite{dowker}
 \beq
\Gamma(\varepsilon)=I-\frac{1}{2}\int_0^\ii dt\
\frac{t^{\varepsilon-1}}{\Gamma(1+\varepsilon)} \Tr
e^{-tL/\mu^2}=I-\frac{1}{2\varepsilon}\ze(\varepsilon|L/\mu^2)\,,
\label{bb} \eeq 
where the zeta function associated with $L$ is defined by  
\beq
\zeta(s|L)=\frac{1}{\Ga(s)}\int_0^\ii dt\ t^{s-1} \Tr e^{-tL}\,,
\hs\hs \ze(s|L/\mu^2)=\mu^{2s}\ze(s|L)\:. \label{mt}
\eeq For a
second order differential operator in 4-dimensions, the integral is
convergent as soon as $\Re s> 2$.

As a consequence,  $\zeta(s|L)$ is  regular at the origin and
one gets the well known result $\zeta(0|L)=A_2(L)$.  This
quantity is  computable (see, for example, \cite{vass03}). 
Furthermore, one may perform a Taylor expansion of the zeta function
\beq 
\zeta(\varepsilon |L)=\zeta(0|L)+\zeta'(0)\varepsilon
+O(\varepsilon^2)\,, 
\eeq 
thus 
\beq 
\Gamma(\varepsilon) =
I-\frac{1}{2\varepsilon}\zeta(0|L)+\frac{\zeta(0|L)}{2}\log \mu^2+
\frac{\ze'(0|L)}{2}+ O(\varepsilon)\,. 
\label{bbb1} \eeq
As a result, one gets the one-loop divergences as well as finite contributions
to the one-loop effective action in terms of the zeta function. With regard to this, a theory is 
one-loop renormalizable as soon as the divergences can be cancelled in a consistent way by the 
renormalization of the bare coupling constants present in the classical action $I$.    

In this paper, we shall investigate modified generalized models, described by a Lagrangian density $F(R,P,Q)$, 
where $R$ is the Ricci scalar, and $P= R_{ij}R^{ij}$, and  
$Q=R_{iijrs}R^{ijrs}$ are quadratic curvature invariants. We do not include   
the Gauss-Bonnet topological invariant because in four dimension, it can be expressed as $G=R^2-4P+Q$.

After some considerations at classical level, the main part of the paper will deal with the one-loop evaluation
of a particular but interesting   $F(R,P,Q)$ model on the de Sitter space, more exactly on its Euclidean version  $S(4)$.

The paper ends with an application to the stability of the de Sitter space within the class of the  
modified gravitational models investigated.

\section{Linear perturbation of $F(R,P,Q)$ model at classical level} 

As warm up exercise, we shall begin with a some considerations at classical level. 
The equation of motion for general $F(R,P,Q)$ model can be found in Ref. \cite{easson} and will not reported here.
In fact, for our purposes, in this Section, it will be sufficient to consider only the trace of the equations of motion, 
which is trivial in Einstein gravity $R=-\kappa^2 T$, but, for a  general $F(R,P,Q)$ model, reads
\begin{equation}
\lap\left(3F'_R+RF'_P \right)+2\nabla_i\nabla_j\left[
   \left(F'_P+2F'_Q\right)R^{ij}\right]-2F+RF'_R
+2\left(PF'_P+QF'_Q\right)=\kappa^2 T\,.
\end{equation}
Requiring $R=R_0$, constant and non negative, $P=P_0$, and $Q=Q_0$ constant, 
one has de Sitter existence condition in vacuum  
\begin{equation}
\left[2F-RF'_R-2PF'_P-2QF'_Q\right]_{R=R_0,P=P_0,Q=Q_0}=0 \,.
\label{deSitterEx}\end{equation}
As a particular but interesting model, let us make the choice
\beq
F(R,P,Q)=f(R)+aP+bQ\,,
\label{f}
\eeq
namely, a generic dependence on $R$, but only linear in the two quadratic invariants $P$ and $Q$.
With regard to this choice, as mentioned in the Introduction, the full  quadratic case
\beq
F(R,P,Q)=R-2\Lambda+cR^2+aP+bQ\,.
\label{2r}
\eeq
namely a Einstein gravity with cosmological constant with the inclusion of curvature square terms is of particular 
interest and  this model has been investigated in many papers, and  was studied in 
the seminal paper \cite{stelle}on the flat space. Note that in this particular case, we may take $b=0$, because the quadratic Gauss-Bonnet invariant $G=R^2-4P+Q$ does not contribute to the equations of motion in four dimensions. Another interesting quadratic model is the Einstein plus conformal invariant quadratic term, i.e.
\beq
F(R,P,Q)=R-2\Lambda+\omega\at\frac{R^2}{3}-2P+Q\ct=R-2\Lambda+\omega C_{ijrs}C^{ijrs}\,,
\label{weyl}\eeq
where $C_{ijrs}$ is the conformal invariant Weyl tensor.  
In the pure conformal case, one has the Weyl conformal gravity, 
and this is a quadratic model admitting exact black hole solutions 
(see for example \cite{mannah,klemm,seba11}).
 
Within the class of modified models (\ref{f}), the dS existence condition becomes  
\begin{equation}
2f_0-R_0f'_{0}=0 \, .
\end{equation}
and the trace equation  in vacuum reads
\beq
\lap(3f'+aR )+2(a+2b)\nabla_i\nabla_j R^{ij}-2f+Rf'=0\,.
\eeq
Making use of contracted Bianchi Identity,
\beq
\nabla_i\nabla_j R^{ij}=\frac12\,\lap R\,,
\eeq
one has
\beq
\lap\left(3f'+2(a+b)R \right)-2f+Rf'=0\,.
\eeq

Perturbing around  dS space, namely   $R=R_0+\delta R $, one arrives at the perturbation equation
\beq
-\lap\delta R+M_0^2\delta R=0\,,
\eeq
in which the scalar degree of freedom  effective mass reads
\beq
M_0^2=\frac{f'_0-R_0f''_0 }{3f''_0+2(a+b)}\,.
\eeq
Thus,  $M_0^2>0$ is a necessary condition for the  stability of the dS solution. 
In these $F(R,P,Q)$ models, besides the massless graviton, 
there exists also a massive spin-two field, as we shall see in the next Section.  

In the particular case of a $F(R,P,Q)=f(R)$ models, there is only the scalaron, and  one recovers the well known condition 
for the dS stability (see, for example \cite{barrow,fara,monica} and references therein)
\beq
\frac{f'_0}{R_0f_0'' }> 1\,.
\eeq

\section{Quantum field fluctuations around the maximally symmetric instantons}

In this Section we will discuss the one-loop quantization of the
model on the a maximally symmetric  space. Of course this should be
considered only an effective approach (see, for instance
\cite{buch}). To start with, we consider the 
Euclidean gravitational model described by the action
\beq
I_E[g]=-\int\:d^4x\,\sqrt{g}\,F(R,P,Q)
=-\int\:d^4x\,\sqrt{g}\,[f(R)+aP+bQ]\,,
\label{action0}\eeq 
with $a,b$ are dimensionless (bare) parameters, 
the Newton constant $G$ being included in the $f(R)$ contribution.
We assume the function $F(R)$ to satisfy the condition (\ref{deSitterEx})
which ensures the existence of constant curvature solutions. This means that
$f(R)$ is not completely arbitrary, but it has to satisfy the equation
\beq
f_0-\frac{R_0f'_0}2=0\,, 
\label{AAA1} 
\eeq 
where here and in the following for the sake of simplicity we use the notation 
$f_0=f(R_0)$, $f'_0=f'(R_0)$ and so on.

We are interested in the dS instanton $S^4$ with positive constant scalar
curvature $R_0$. This is a maximally symmetric space
having covariant conserved curvature tensors. Its  
metric may be written in the form
\beq
ds_E^2=d\tau^2(1-H_0^2r^2)+\frac{dr^2}{(1-H_0^2r^2)}+r^2dS_2^2\,,
\eeq
$dS^2$ being the metric of the two-dimensional sphere $S^2$.
The finite volume reads
\beq
V(S^4)=\frac{384\pi^2}{R_0^2}\,,\qquad\qquad R_0=12H^2_0\,,
\eeq
while Riemann and Ricci tensors are given by
\beq 
R^{(0)}_{ijrs}=\frac{R_0}{12}\at g^{(0)}_{ir}g^{(0)}_{js}-
g^{(0)}_{is}g^{(0)}_{jr}\ct \:, \hs
R^{(0)}_{ij}=\frac{R_0}{4}\,g^{(0)}_{ij}\,. 
\label{AAA2} 
\eeq 

Now let us consider small fluctuations around the maximally
symmetric instanton. In the action (\ref{action0}) then we set
\beq 
g_{ij}\longrightarrow g_{ij}+h_{ij}\:,\hs
g^{ij}\longrightarrow g^{ij}-h^{ij}+h^{ik}h^j_k+{\cal O}(h^3)\:,\hs
h=g^{ij}h_{ij}\:, 
\eeq 
where from now on $g_{ij}\equiv g^{(0)}_{ij}$ is the metric of the
maximally symmetric space and 
as usual, indices are lowered and raised
by the means of such a metric. 
Up to second order in $h_{ij}$ one has 
\beq
\sqrt{g}\longrightarrow \sqrt{g}\aq 1+\frac12h+\frac18h^2-\frac14h_{ij}h^{ij}+{\cal O}(h^3)\cq
\eeq 
and 
\beq
 R &\sim& R_0-\frac{R_0}{4}\,h+\nabla_i\nabla_jh^{ij}-\lap h
\nn \\ && +\frac{R_0}{4}\,h^{jk}h_{jk} -\frac14\,\nabla_ih\nabla^ih
-\frac14\,\nabla_kh_{ij} \nabla^kh^{ij} +\nabla_ih^i_k\nabla_jh^{jk}
-\frac12\,\nabla_jh_{ik}\nabla^ih^{jk} \:, 
\eeq 
where $\nabla_k$ represents the covariant derivative in the unperturbed metric
$g_{ij}$.
More complicated expressions are obtained for the other invariants $P,Q$,
but for our aim it is not necessary to write them explicitly. 

By performing a Taylor expansion of the Lagrangian around de Sitter metric,
up to second order in $h_{ij}$, we get 
\beq 
I_E[g]\sim-\int\:d^4x\,\sqrt{g}\:
\aq F(R_0,P_0,Q_0)+\frac{hX}2+{\cal L}_2\,\cq\,, 
\label{AAA3} 
\eeq 
where ${\cal L}_2$ represents the second-order contribution and
$X=f_0-R_0f'_0/2$ vanishes when $f(R)$ satisfies the de Sitter existence solution (\ref{AAA1}).

It is convenient to carry out the standard expansion of the tensor
field $h_{ij}$ in irreducible components \cite{frad}, namely 
\beq
h_{ij}=\hat
h_{ij}+\nabla_i\xi_j+\nabla_j\xi_i+\nabla_i\nabla_j\sigma
+\frac14\,g_{ij}(h-\lap_0\sigma)\:, 
\label{tt}\eeq 
where $\si$ is
the scalar component, while $\xi_i$ and $\hat h_{ij}$ are the vector
and tensor components with the properties 
\beq
\nabla_i\xi^i=0\:,\hs\hs \nabla_i\hat h^{ij}=0\:,\hs\hs \hat h_i^i=0\:. 
\label{AAA4} \eeq 
In terms of the irreducible components
of the $h_{ij}$ field, 
the Lagrangian density, disregarding total derivatives, becomes
\begin{eqnarray}
{\cal L}_2&=&{\cal L}_{hh}+2\,{\cal L}_{h\si}+{\cal L}_{\si\si}+{\cal L}_V+{\cal L}_T\,,
\end{eqnarray}
where ${\cal L}_{hh},{\cal L}_{h\si},{\cal L}_{\si\si}$ represent the scalar 
contribution (a $2\X2$ matrix),
while ${\cal L}_V$ and ${\cal L}_T$ represent the vector and tensor contributions respectively.
One has
\begin{eqnarray}
{\cal L}_{hh}&=&h\,\aq
\frac{9{f''_0}{\Delta}^2}{32}
-\frac{3{f'_0}{\Delta}}{32}
+\frac{aR_0{\Delta}}{16}
+\frac{bR_0{\Delta}}{16}
+\frac{3{f''_0}R_0{\Delta}}{16}
+\frac{{f''_0}R_0^2}{32}
-\frac{{f'_0}R_0}{32}
\cp\nonumber\\&&\ap\hs\hs
+\frac{X}{16}
+\frac{3a{\Delta}^2}{16}
+\frac{3b{\Delta}^2}{16}
\cq\, h\:,
\end{eqnarray}

\begin{eqnarray}
{\cal L}_{h\si}&=&h\,\aq
-\frac{9{f''_0}{\Delta}^3}{32}
+\frac{3{f'_0}{\Delta}^2}{32}
-\frac{3}{16}{f''_0}R_0{\Delta}^2
-\frac{1}{32}{f''_0}R_0^2{\Delta}
+\frac{{f'_0}R_0{\Delta}}{32}
\cp\nonumber\\&&\ap\hs\hs
-\frac{3a{\Delta}^3}{16}
-\frac{3b{\Delta}^3}{16}
-\frac{1}{16}aR_0{\Delta}^2
-\frac{1}{16}bR_0{\Delta}^2\cq\,\sigma\,,
\end{eqnarray}

\begin{eqnarray}
{\cal L}_{\si\si}&=&\sigma\,\aq
\frac{9{f''_0}{\Delta}^4}{32}
-\frac{3{f'_0}{\Delta}^3}{32}
+\frac{3}{16}{f''_0}R_0{\Delta}^3
+\frac{1}{32}{f''_0}R_0^2{\Delta}^2
-\frac{1}{32}{f'_0}R_0{\Delta}^2
\cp\nonumber\\&&\ap\hs\hs
-\frac{3X{\Delta}^2}{16}
-\frac{R_0X{\Delta}}{16}
\frac{3a{\Delta}^4}{16}
+\frac{3b{\Delta}^4}{16}
+\frac{1}{16}aR_0{\Delta}^3
+\frac{1}{16}bR_0{\Delta}^3\cq\,\sigma\,,
\end{eqnarray}

\begin{eqnarray}
{\cal L}_V&=&\xi^k\,\aq\frac{1}{8}XR_0+\frac{1}{2}X\Delta\cq\,\xi_k\:,
\end{eqnarray}

\begin{eqnarray}
{\cal L}_T&=&\hat h^{ij}\,\aq
+\frac{{f'_0}{\Delta}}{4}
-\frac{{f'_0}R_0}{24}
-\frac{X}{4}
\frac{a{\Delta}^2}{4}
+b{\Delta}^2
+\frac{aR_0{\Delta}}{24}
-\frac{bR_0{\Delta}}{3}
-\frac{aR_0^2}{72}
+\frac{bR_0^2}{36}\cq\,\hat h_{ij}\,.
\end{eqnarray}
where $\lap=g^{ij}\nabla_i\nabla_j$ is the Laplace-Beltrami operator in the unperturbed 
metric $g_{ij}$, which is 
a solution of field equations, but only if $X=0$.
We have written the above expansions around a maximally symmetric space
which in principle could not be a solution. This means that the function 
$f(R)$ can be arbitrary.

As it is well known, invariance under diffeormorphisms renders the
operator in the $(h,\si)$ sector not invertible. One needs a gauge
fixing term and a corresponding ghost compensating term. Here we choose
the harmonic gauge, that is
\beq
\chi_j=-\nabla_ih^i_j-\frac12\,\nabla_j h=0\,,
\eeq
and the gauge fixing term
\beq 
{\cal L}_{gf}=\frac12\,\chi^iG_{ij}\chi^j\,,\hs\hs G_{ij}=\ga\,g_{ij}\,.
\label{AAA5} 
\eeq
The corresponding ghost Lagrangian reads \cite{buch} 
\beq
{\cal L}_{gh}= B^i\,G_{ik}\frac{\de\,\chi^k}{\de\,\ep^j}C^j\,,
\label{AAA6} \eeq 
where $C_k$ and $B_k$ are the ghost and anti-ghost
vector fields respectively, while $\de\,\chi^k$ is the variation of
the gauge condition due to an infinitesimal gauge transformation of
the field. In this case it reads
 \beq
\de\,h_{ij}=\nabla_i\ep_j+\nabla_j\ep_i\segue
\frac{\de\,\chi^i}{\de\,\ep^j}=g_{ij}\,\lap+R_{ij}\,.
\label{AAA7} \eeq 
Neglecting total derivatives, one has 
\beq {\cal L}_{gh}=B^k\,\ga\,\at\lap+\frac{R_0}{4}\ct\,C_k\,.
\label{AAA8} \eeq 
In irreducible components one finally obtains
\begin{eqnarray}
{\cal L}_{gf} &=&\frac{\ga}2\aq\xi^k\,\at\lap_1+\frac{R_0}4\ct^2\,\xi_k
    +\frac{3\rho}{8}\,h\,\at\lap_0+\frac{R_0}3\ct\,\lap_0\,\si
\cp\nn\\&&\hs\ap
    -\frac{\rho^2}{16}\,h\,\lap_0\,h
-\frac{9}{16}\,\si\,\at\lap_0+\frac{R_0}3\ct^2\,\lap_0\,\si\cq
\label{AAA10} \eeq 
\beq {\cal L}_{gh} &=&
\ga\aq\hat B^k\at\lap_1+\frac{R_0}{4}\ct\hat C_k
+\frac{\rho-3}{2}\,\hat b\,\at\lap_0-\frac{R_0}{\rho-3}\ct\,\lap_0\hat c\cq\,,
\eeq
where ghost irreducible components are defined by
\beq
C_k&=&\hat C_k+\nabla_k\hat c\,,\hs\hs \nabla_k\hat C^k=0\,,
\nn\\
B_k&=&\hat B_k+\nabla_k\hat b\,,\hs\hs \nabla_k\hat B^k=0\,.
\label{AAA11} \eeq 

\section{One-loop effective action}

In order to compute the one-loop contributions to
the effective action one has to consider the path integral for the
bilinear part 
${\cal L}= {\cal L}_2+\,{\cal L}_{gf}+{\cal L}_{gh} $
of the total Lagrangian and take into
account the Jacobian due to the change of variables with respect to
the original ones. In this way one gets  \cite{frad,buch} 
\beq
Z^{(1)}&=&\at\det G_{ij}\ct^{-1/2}\,\int\,D[h_{ij}]D[C_k]D[B^k]\:
\exp\,\at -\int\,d^4x\,\sqrt{g}\,{\cal L}\ct
\nn\\
&=&\at\det G_{ij}\ct^{-1/2}\,\det J_1^{-1}\,\det J_2^{1/2}\,
\nn\\&&\times \int\,D[h]D[\hat h_{ij}]D[\xi^j]D[\si] 
D[\hat C_k]D[\hat B^k]D[c]D[b]\:\exp\, \at-\int\,d^4x\,\sqrt{g}\,{\cal L}\ct\,, 
\eeq where $J_1$  and $J_2$ are the Jacobians due to the
change of variables in the ghost and tensor sectors respectively
\cite{frad}. They read 
\beq J_1=\lap_0\,,\hs\hs J_2=\at\lap_1+\frac{R_0}{4}\ct\at\lap_0+\frac{R_0}{3}\ct\,\lap_0\,,
\label{AAA13} \eeq 
and the determinant of the operator $G_{ij}$ in this case is trivial.
Due to the presence of curvature, the Euclidean gravitational action
is not bounded from below, because  arbitrary negative contributions 
can be induced on $R$, by conformal 
rescaling of the metric. For this reason, 
we have also used the Hawking prescription of integrating over
imaginary scalar fields. Furthermore, the problem of the presence of
additional zero modes introduced by the decomposition (\ref{tt}) can
be treated  making use of the method presented in Ref.~\cite{frad}.

Now, a straightforward computation leads to the following off-shell
one-loop contribution to the ``partition function'' 
\beq
e^{-\Ga^{(1)}}\equiv Z^{(1)}&=&
\ag\det\aq\at\lap_1+\frac{R_0}4+\frac{X}{\ga}\ct\,
\at\lap_0+\frac{R_0}2+\frac{X}{\ga}\ct\,L_0^+\,L_0^-\,L_2^+\,L_2^-\cq\cg^{-1/2}
\nn\\&&\hs\hs\X\:
\det\aq\at\lap_0+\frac{R_0}2\ct\,\at\lap_1+\frac{R_0}4\ct\cq\,,
\label{PF}\eeq
where
\beq
L_0^\pm&=&\lap_0-\frac{2f'_0-5R_0f''_0-2R_0(a+b)}{4(3f''_0+2(a+b)]}
\nn\\&&\hs\pm
   \frac{\sqrt{[2f'_0-5R_0f''_0-2R_0(a+b)]^2-8[3f''_0+2(a+b)][2X-R_0(f'_0-R_0f''_0)]}}
{4(3f''_0+2(a+b)]}\,,
\nn\eeq
\beq
L_2^\pm&=&\lap_2+\frac{6f'_0+R_0(a-8b)}{ 12 (a + 4 b)}
    \pm\frac{\sqrt{(2f'_0+aR_0)^2+ 16X(a+4b)}}{4(a + 4 b)}\,,
\nn\eeq
$\lap_0,\lap_1,\lap_2$ being respectively the Laplacian operators acting on
scalars, transverse vectors and transverse, traceless tensors and of course
$a+4b\neq0$.

The partition function in (\ref{PF}) explicitly depends on the gauge parameter $\ga$, 
but it is known that when one goes ``on-shell'', that is when one imposes the background metric
$g_{ij}$ to be a solution of the field equation, 
the one-loop partition function becomes gauge independent. 
In our case we have simply to perform the limit $X\to0$ obtaining
\beq
Z^{(1)}_{on-shell}&=&
\ag\det\aq\at-\lap_0+\frac{f'_0- R_0f''_0}{3f''_0+2(a+b)}\ct\cq\cg^{-1/2}\,
\ag\det\at-\lap_1-\frac{R_0}4\ct\cg^{1/2}
\nn\\&&\hs\X
\ag\det\aq\at-\lap_2+\frac{R_0}6\ct\,
\at-\lap_2+\frac{(2b-a)R_0-3f'_0}{3(a+4b)}\ct\cq\cg^{-1/2}\,.
\label{PFonS}\eeq

As a consequence, the on-shell one-loop effective action reads 
\beq
\Ga_{on-shell}=I_E(g)+\Ga^{(1)}_{on-shell}\,,
\nn\eeq
\beq
\Ga^{(1)}_{on-shell}&=&\frac12\,\log\det\at\frac1{\mu^2}\,\aq-\lap_0+\frac{f'_0-R_0f''_0}{3f''_0+2(a+b)}\cq\ct
\nn\\&&\hs
  -\frac12\,\log\det\at\frac1{\mu^2}\,\aq-\lap_1-\frac{R_0}4\cq\ct
\nn\\&&\hs\hs
  +\frac12\,\log\det\at\frac1{\mu^2}\,\aq-\lap_2+\frac{R_0}6\cq\ct
\label{1-loopEA}\\&&\hs\hs\hs
  +\frac12\,\log\det\at\frac1{\mu^2}\,\aq-\lap_2+\frac{(2b-a)R_0-3f'_0}{3(a+4b)}\cq\ct\,.
\nn\eeq
As usual an arbitrary renormalization parameter $1/\mu^2$ has
been introduced for dimensional reasons.

The one-loop contribution $\Ga^{(1)}$ to the effective action can be computed by 
by using zeta-function techniques. The eigenvalues of Laplacian operators
on $S^4$ are explicitly known and so the determinant of all operators appearing
in (\ref{1-loopEA}) in principle can be calculated. We refer the interested reader 
to Ref.~\cite{cogno}, where all details of computation can be found.

\section{Discussion and conclusions}

We conclude the paper with several remarks.
First, the one-loop effective action result is in agreement 
with a similar one given in \cite{cogno8} and
in the limit $a\to0,b\to0$ becomes identical to those in \cite{cogno}, 
where only $f(R)$ modified gravity has been considered. 

Both equations in (\ref{PF}) and (\ref{PFonS}) have been derived by 
assuming $a+4b\neq0$. 
As a useful check, we note that when $a=-4b$, the original classical Lagrangian 
density can be written in the form
\beq
f(R)+aP+bQ=\tilde f(R)+bG\,,\hs\hs \tilde f(R)=f(R)-bR^2\,,\hs G=R^2-4P+Q\,,
\nn\eeq
where $G$ is the Gauss-Bonnet topological invariant which
does not contribute to the field equations. 
In such a case, as expected, the one-loop contribution to the partition function becomes
\beq
\tilde Z^{(1)}_{on-shell}&=&
\ag\det\aq\at-\lap_0+\frac{\tilde f'_0- R_0\tilde f''_0}{3\tilde f''_0}\ct\cq\cg^{-1/2}
\nn\\&&\hs\hs\X
\ag\det\at-\lap_1-\frac{R_0}4\ct\cg^{1/2}
\ag\det\at-\lap_2+\frac{R_0}6\ct\cg^{-1/2}\,,
\nn\eeq
which again is the result obtained for a pure $\tilde f(R)$ modified gravity \cite{cogno}.

Another interesting particular case is the one in which the Einstein-Hilbert 
Lagrangian density is modified by a term proportional to the the square of the Weyl tensor.
Such a classical model has a de Sitter solution  and
in the absence of Einstein-Hilbert term, the on-shell, one-loop
contribution to the partition function trivially vanishes, but if
$a=-4b,b=\om$ and, as in (\ref{weyl})
\beq
f(R)+aP+bQ=\tilde f(R)+\om\at\frac{R^2}3-2P+Q\ct\,,
\hs\hs \tilde f(R)=\frac{R}{16\pi G_N}=M_P^2R\,,
\nn\eeq
$G_N,M_P$ being respctively the Newton constant and Planck mass,
the on-shell one-loop contribution reads
\beq
\tilde Z^{(1)}_{Weyl}=\ag\det\at-\lap_1-\frac{R_0}4\ct\cg^{1/2}
\ag\det\aq\at\-\lap_2+\frac{R_0}6\ct\,\at-\lap_2+\frac{R_0}3
    -\frac{\tilde f'_0}{2\om}\ct\cq\cg^{-1/2}\,.
\nn\eeq
In contrast with previous cases, here  the contribution due to scalar components 
is vanishing. 

As an important application we discuss the stability of de Sitter space.  
To this aim we have to recall that the eigenvalues $\la_n$ of Laplacian-Beltrami 
$-\lap$ operators on $S^4$ have the form
\beq
\la_n=\frac{R_0}{12}\,\aq(n+\nu)^2-\al\cq\,,\hs
     g_n=c_1(n+\nu)+c_3(n+\nu)^3\,,\hs n=0,1,2,...
\nn\eeq
$g_n$ being the corresponding degeneracy and $\nu,c_1,c_3$ being dimensionless quantities,
which depend on the operator one is dealing with. In particular one has
\beq
\begin{array}{llll}
-\lap_0\segue\nu=\frac32\,,&\hs\al=\frac94\,,&\hs c_1=-\frac1{12}\,,&\hs c_3=\frac13\,,\\
-\lap_1\segue\nu=\frac52\,,&\hs\al=\frac{13}4\,,&\hs c_1=-\frac94\,,&\hs c_3=1\,,\\
-\lap_2\segue\nu=\frac72\,,&\hs\al=\frac{17}4\,,&\hs c_1=-\frac{125}{12}\,,&\hs c_3=\frac53\,.
\end{array}
\eeq
We see that only the scalar Laplacian $-\lap_0$ has a null eigenvalue,
while the minimum eigenvalue of $-\lap_1$ is $R_0/4$ and 
the minimum eigenvalue of $-\lap_2$ is $2R_0/3$.

In Eq. (\ref{1-loopEA}) we are dealing with operators of the kind $L=-\lap+M^2$ and so,
in order to have stability of de Sitter solution, we have to assume all
eigenvalues of $L$ to be positive. In this way we obtain restrictions on 
the function $F(R,P,Q)$. 

Looking at (\ref{1-loopEA}) we see that, 
independently on the classical action, a zero-mode is present coming 
from the Laplacian-like operator $-\lap_1-R_0/4$. 
In principle, other zero modes may be present and all of these 
can be treated according to Ref.~\cite{frad}.

The other operators in (\ref{1-loopEA})  which could have vanishing or negative
eigenvalues are 
\beq
L_0&=&-\lap_0+M_0^2\,,\hs\hs M_0^2=\frac{f'_0-R_0f''_0}{3f''_0+2(a+b)}\,,
\nn\\
L_2&=&-\lap_2+M_2^2\,,\hs\hs M_2^2=\frac{R_0(2b-a)-3f'_0}{3(a+4b)}\,,
\nn\eeq 
but in the case in which
\beq\ag\begin{array}{lll}
M_0^2>0&\segue&\frac{f'_0-R_0f''_0}{3f''_0+2(a+b)}>0\,,\\
M_2^2+\frac23\,R_0>0&\segue&\frac{(a+10b)R_0-3f'_0}{3(a+4b)}>0\,.
\end{array}\cp\eeq
In the particular cases in which $M^2_0=0$ and/or $M^2_2=-2/3R_0$ there are other
zero-modes which have to be treated as the previous ones. 

In the interesting case $f(R)=M^2_P(R-2\Lambda)$, linear in the curvature, 
the dS stability conditions become
\beq\ag\begin{array}{l}
M_0^2=\frac{M^2_P}{2(a+b)}\,,\\
M_2^2=\frac{4\La(a+10b)-3M^2_P}{3(a+4b)}>0\,,
\end{array}\cp\quad
\Longrightarrow
\ag\begin{array}{l}
a+b>0\\
a+4b>0\\
a+10b>\frac{3M^2_P}{4\La}
\end{array}\cp\,,
\qquad
\ag\begin{array}{l}
a+b>0\\
a+4b<0\\
a+10b<\frac{3M^2_P}{4\La}
\end{array}\cp\,.
\eeq
We see that depending on the arbitrary parameters $a,b$ the solution can be 
stable or unstable and in order to have a stable solution at least one of
the two parameters has to be positive. In particular,
in the special cases $a=0$ or $b=0$ one gets the stability conditions
\beq
a>\frac{3M^2_P}{4\La}>0\,,&&\hs b=0\,,
\nn\\
b>\frac{3M^2_P}{40\La}>0\,,&&\hs a=0\,.
\nn\eeq

In summary, here we have evaluated the one-loop effective action for 
a specific modified gravity model in de Sitter space. 
Generalized zeta regularization could be used to obtain a finite answer for the 
functional determinants in the effective action, what has proven to 
be a very convenient procedure. 

The important lesson to be drawn 
from this calculation, generalizing the previous 
program for one-loop Einstein gravity and $f(R)$ modified gravity 
in the de Sitter background is that quantum corrections may to 
destabilize the classical de Sitter universe, as we have explicitly 
verified in the examples.

\ack{This article is dedicated to  Professor Stuart Dowker. His seminal work on zeta function regularization
has been a continue inspiration for many researchers. We would like to thank the organizers for the kind invitation to present a contribution to the Special Issue in his honor.}


\end{document}